\documentclass[prl,10pt,twocolumn,notitlepage,floatfix,footinbib,superscriptaddress,showpacs,showkeys,aps]{revtex4-1}

\usepackage{graphicx}
\usepackage[usenames,dvipsnames]{color}
\usepackage{hyperref}
\usepackage{amsmath}

\newcommand{\bs}[1]{\boldsymbol{#1}}

\begin{document}
\title{Interaction-tuned Anderson versus Mott localization}
\author{Andrey~E.~Antipov}\email{aantipov@umich.edu}
\affiliation{Department of Physics, University of Michigan, Ann Arbor, Michigan 48109, USA}
\author{Younes Javanmard}
\affiliation{Max-Planck-Institut f\"{u}r Physik komplexer Systeme, 01187 Dresden, Germany}
\author{Pedro Ribeiro}\email{ribeiro.pedro@gmail.com}
\affiliation{CeFEMA, Instituto Superior T\'{e}cnico, Universidade de Lisboa, Av. Rovisco Pais, 1049-001 Lisboa, Portugal}
\affiliation{Russian Quantum Center, 143025, Skolkovo, Moscow region, Russia}
\author{Stefan Kirchner}\email{stefan.kirchner@correlated-matter.com}
\affiliation{Center for Correlated Matter, Zhejiang University, Hangzhou,  Zhejiang 310058, China}

\date{\today}

\begin{abstract}
Disorder or sufficiently strong interactions can render a metallic state unstable causing it to turn into an insulating one. Despite the fact that the interplay of these two routes to a vanishing conductivity has been a central research topic, a unifying picture has not emerged so far. Here, we establish that the two-dimensional Falicov-Kimball model, one of the simplest lattice models of strong electron correlation does allow for the study of this interplay. In particular, we show that this model at particle-hole symmetry possesses three distinct thermodynamic insulating phases and exhibits Anderson localization. The previously reported metallic phase is identified as a finite-size feature due to the presence of weak localization. We characterize these phases by their electronic density of states, staggered occupation, conductivity, and the generalized inverse participation ratio. The implications of our findings for other strongly correlated systems are discussed.  
\end{abstract}


\maketitle
A Fermi liquid becomes unstable in the presence of strong disorder or Coulomb repulsion. Although this has been known for a long time, the interplay of disorder and electron-electron interaction near the metal-insulator transition is still at the forefront of condensed matter research~\cite{Belitz1994,Basko2006}. This is at least in part due to the lack of general techniques  to tackle strong electron interactions in the presence of  disorder in higher-dimensional systems. 
As is well known, a metal is a good electrical and thermal conductor, as defined by a non-vanishing value of the DC conductivity. 
An insulator can therefore be defined as a system for which this quantity vanishes.  
In systems where the electron-electron interaction can be neglected, two types of insulators, {\it i.e.} band insulators and Anderson insulators, can exist. In the presence of interactions, the situation is richer, as {\it e.g.} Mott insulators, excitonic insulators, or even Wigner crystals may form. Under the special condition of perfect nesting, an ordered  band insulator-like  state is also possible as a result of an electronic phase transition.
Following the seminal work of Basko et al.~\cite{Basko2006}, the many-body localized state, {\it i.e.} the insulating state that has its origin in the interplay of disorder and interaction, has recently received increased attention \cite{Nandkishore2015,BarLev2015}. 

Typically, the term disorder is understood as being synonymous to quenched disorder, {\it e.g.} when random variables are assumed not to evolve with time. The average over the disorder is taken to mimic the spatial self-averaging of the system. This is in contrast to the annealed disorder where the disorder follows a thermal distribution. When (quantum) dynamics is neglected the partition function of any system can appear as that of an annealed disorder problem. In systems where a separation of time scales permits this neglect for the slow fields, occurrence of a localization without explicit disorder may be possible~\cite{Gavish2005,Yao2014,Schiulaz2015,Papic2015}. Only at sufficiently high temperature, where every configuration carries essentially the same thermal weight, the difference between quenched and annealed disorder is immaterial. 

The Falicov-Kimball (FK) model ~\cite{Falicov1969} is one of the simplest lattice models of interacting electrons. It was originally developed to describe the metal-insulator transition in the context of f-electron systems and can be understood as a limiting case of the Hubbard model where the dynamics of one of the spin-degenerate fermion species is neglected. Thus, these fermions become immobile. 
Therefore, the partition function of the FK model can be seen as one of annealed disorder of local $f$-electron occupation numbers, allowing for the possibility to observe the Anderson localization in the absence of explicit disorder. At half-filling, the FK model describes a charge-ordered state below some $U$-dependent transition temperature $T_c(U)$ at all non-vanishing values of the interaction strength $U$ between localized and itinerant electrons \cite{Brandt1986,Kennedy1986,Maska2006,Zonda2009}. This charge-ordered state is commonly referred to as a charge-density wave (CDW) state.
Within the standard approach to strongly correlated electron systems, {\it i.e.} the dynamical mean field theory (DMFT), the resulting effective impurity action, associated with the FK model can be solved exactly~\cite{Metzner1989, Brandt1986, Brandt1990, VanDongen1992,Georges1996,Kotliar2006}. For this reason, the FK model is often taken as a test bed for DMFT approaches and its extensions~\cite{Freericks2003,Antipov2014,Ribic2016}.

\begin{figure}[ht]\begin{center}
\includegraphics[width=1.0\columnwidth]{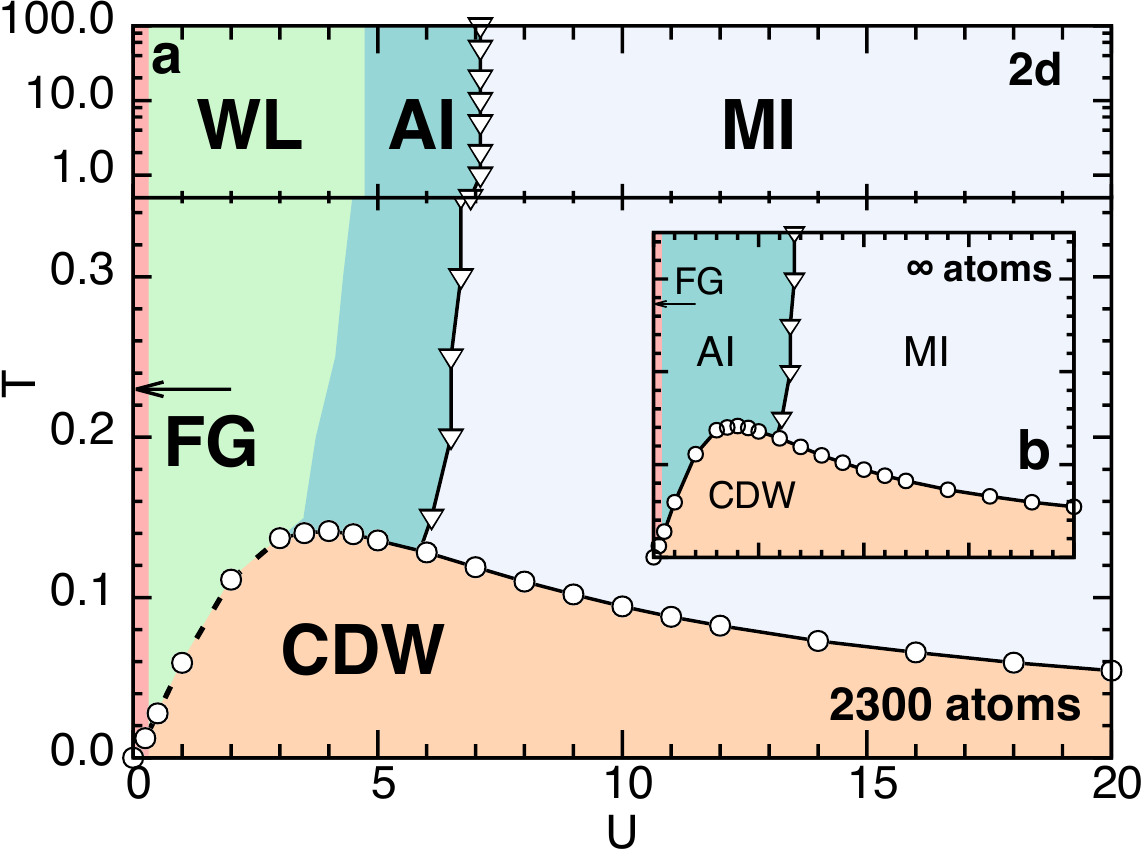}\end{center}\vspace{-1em}
\caption{Phase diagram of the particle-hole symmetric FK model in $2$d in $U$-$T$ plane, obtained by lattice Monte-Carlo, consisting of different phases: Fermi gas (FG) at $U=0$, charge-density wave insulator (CDW) at low temperature and all non-zero values of $U$.  High temperature phases: Anderson insulator (AI) at intermediate values of $U$ crossing over to a weakly localized (WL) at smaller $U$, Mott-like insulator (MI) at large $U$. The points and lines show phase boundaries, the dashed line indicates the first order phase transition between WL and CDW phases. Inset: extrapolation to the thermodynamic limit.}
\label{fig1}
\end{figure}

In this letter we revisit the finite temperature phase diagram of the two-dimensional FK model at half-filling using state-of-the-art Lattice Monte Carlo techniques. 
The main results are summarized in Fig.~\ref{fig1} that shows the phase diagram of the particle-hole symmetric FK model in the interaction $U$ - temperature $T$ plane. By analyzing different observables of the mobile electrons we found multiple different regions with qualitatively distinct properties: a CDW at low temperatures, at high temperature - large U Mott-like insulator (MI) phase,  a non-interacting Fermi gas (FG) at $U=0$, and, a central result of the paper, a region overlooked in previous studies \cite{Maska2006} where weak localization (WL) induces a finite-volume crossover between a  bad-metal (see below) and an Anderson-insulator (AI). 

The Hamiltonian of the FK-model is 
\begin{equation}
H= -t \sum_{\left\langle ij \right\rangle} c^\dagger_{i} c_{j} - \mu
\sum_{i}  \left(c^\dagger_{i} c_{i} + n_{f,i}\right) + \sum_i U c^\dagger_{i} c_{i}  n_{f,i},
\end{equation} 
where $c^\dagger_{i}$ creates a $c$-electron and $n_{f,i}$ counts the number of $f$-electrons on site $i$; $\mu$ is the chemical potential of the system. In the following we set $\mu=U/2$ corresponding to the half-filling condition for both species; $U$ is an on-site interaction potential, and $t$ is the hopping strength of $c$-electrons used as a unit of energy $t=1$. The summation $\sum_{\left\langle i,j \right\rangle}$ runs over all nearest-neighbor pairs on a square lattice with a volume $V=L^2$. 
Since $n_{f,i}$ is a conserved quantity, the partition function of the model is given by a summation over non-interacting contributions from every configuration $n_f$ of f-electrons $Z=\sum_{n_f} e^{- \beta E[n_f]  }$ where 
$ E[n_f] = - \mu \sum_{i} n_{f,i} -\beta^{-1} \ln \det \left[ 1 + e^{- \beta \bs{h}} \right] $ and $\bs{h}_{i,j}[n_f] = - t \delta_{\left|i-j\right|,1} +U n_{f,i} \delta_{i,j}$ a matrix of dimensions $V \times V$. 
In this form, the partition function is amenable to evaluation by a Lattice Monte Carlo sampling. 
The configurations of the classical random variables   $n_{f,i}=0,1$ are sampled with a Boltzmann factor and thus the 
 partition function $Z$ of the model is equivalent to that of a sign-problem free annealed disorder problem. The sampling using Chebyshev polynomials is used to study the properties of the system at large $V \geq 32^2$ \cite{Motome1999}.

\begin{figure}[t]
\begin{center}
{\includegraphics[width=\columnwidth]{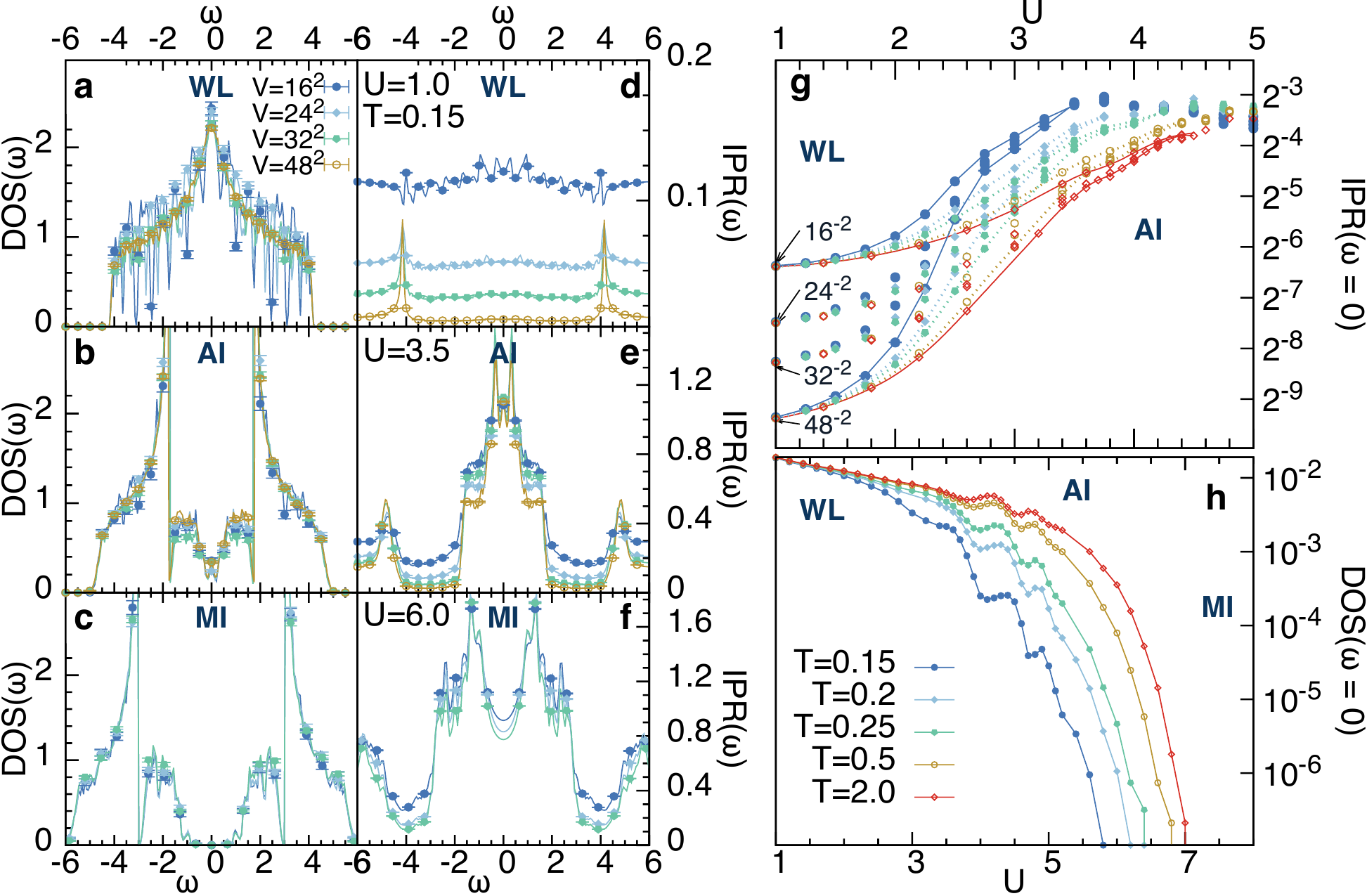}}
\end{center}\vspace{-1em}
\caption{
\textbf{a}-\textbf{c}, DOS and \textbf{d}-\textbf{f}, IPR as a function of frequency $\omega$ at $U = 1, 3.5, 5.25$, $T = 0.15$ and system sizes $16^2, 24^2, 32^2, 48^2$. \textbf{g}, IPR and  \textbf{h}, DOS at $\omega = 0$ as a function of $U$ at different temperatures and system sizes. $\mathrm{DOS}(0)$ is independent of system size.} 
\label{fig2}
\end{figure}

The staggered charge susceptibility of $f$-electrons and the specific heat of the two-dimensional problem were studied in \citep{Maska2006,Maska2005} and the properties of the CDW phase have been established.
Here, we mainly focus on the behavior of the $c$ electrons. For  fixed $n_f$ configuration, we sample the $c$-electron density of states (DOS), $\text{DOS}(\omega) = \sum_n \delta(\omega- \varepsilon_n)$, and
 introduce the energy resolved inverse participation ratio (IPR): $\text{IPR}_{n_f}(\omega) = {\text{DOS}(\omega)}^{-1}\sum_n\sum_i \delta(\omega- \varepsilon_n)  \psi_{n,i}^4$, where $\psi_{n,i}$ is the $n$-th eigenfunction of $\bs h[n_f]$ with energy  $\varepsilon_n$ at site $i$ \footnote{The delta-peaks $\delta(\omega - \varepsilon_n)$ are sampled as Lorentzian functions with $\xi=10^{-3}$ broadening, shown results are independent of this broadening}. The IPR is an ideal measure of the degree of localization of $c$-electron single particle states \cite{Ferdinand2008}. For a completely itinerant system, the IPR scales with the inverse of $V$ while $\text{IPR}\to \mathrm{const}$ in the fully localized limit or when the system is gapped \footnote{IPR at frequency $\omega$ can also be obtained by binning within a fixed energy window around frequency $\omega$, resulting in the same quantity as defined in the text whenever $DOS(\omega) \neq 0$, and otherwise $\text{IPR}=0$ (not shown)}.
Averaging over $n_f$, we obtain  $\text{IPR}(\omega)=\sum_{n_f} e^{- \beta E[n_f]  }\text{IPR}_{n_f}(\omega) /Z$  that quantifies the average degree of localization at energy $\omega$.  As we will show, the $\text{IPR}(\omega)$ and $\text{DOS}(\omega)$ provide a convenient characterization of the evolution from a WL region at small $U$ and finite $L$ to the AI and MI phases at stronger interaction. The results for DOS and IPR are shown in Fig. \ref{fig2}. 
 
\begin{figure}[t]
\begin{center}
{\vspace{-1em}\includegraphics[width=\columnwidth]{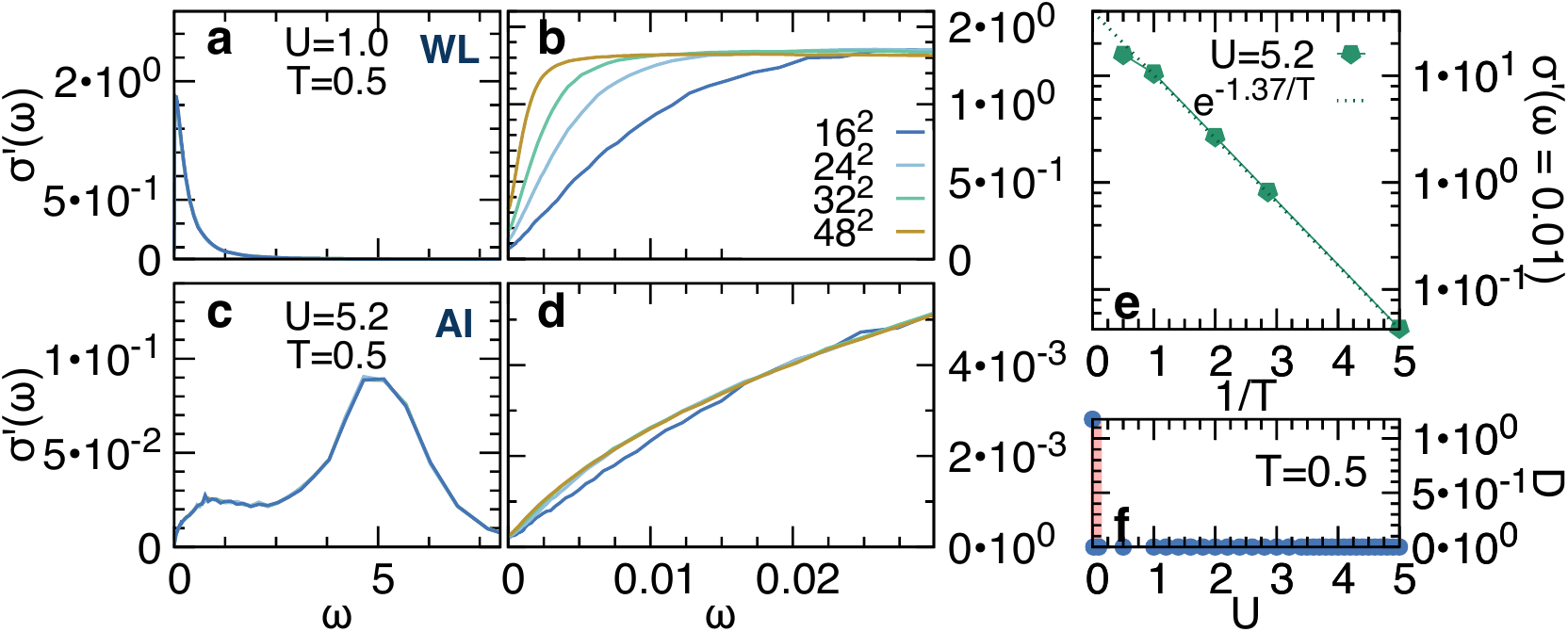}}
\end{center}\vspace{-1em}
\caption{\textbf{a,b,c,d,} Real part of the AC conductivity $\sigma'(\omega)$ as a function of frequency $\omega$ at $T=0.5$, $U=1.0$ (a, b), $5.2$ (c,d). Panels (b),(d) show $\sigma'(\omega$) for $|\omega| \leq 0.025$ at different system sizes.  \textbf{e}, The temperature dependence of $\sigma'$ at $\omega = 0.01$ and $U=5.2$. \textbf{f}, Drude weight $D$ as a function of $U$ at $T=0.5$.
}
\label{fig3}
\end{figure}

\begin{figure}[ht]
\begin{center}
{\vspace{-1em}\includegraphics[width=\columnwidth]{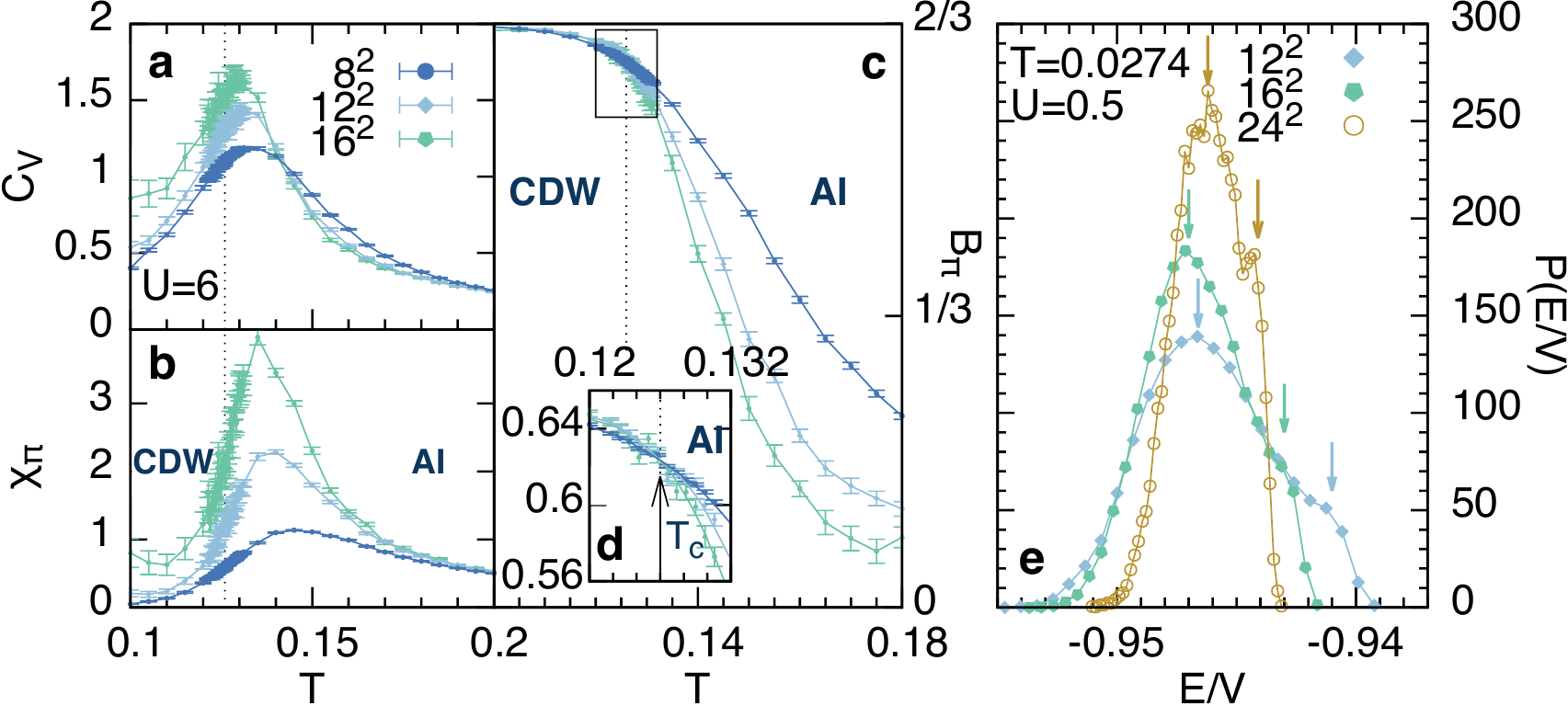}}
\end{center}\vspace{-1em}
\caption{\textbf{a,b,c,} Specific heat, $f$-electron susceptibility and Binder cumulant for momentum $q=\{\pi,\pi\}$ at $U/t = 6$ and volumes $8^2, 12^2, 16^2$, illustrating the phase transition to the CDW state. Dashed arrow is the transition temperature $T_{\mathrm{CDW}}$ obtained by the crossing of the Binder-cumulant curves, shown in Fig. \ref{fig4}d. \textbf{e,} The distribution of energy values at the proximity of transition temperature to CDW-state at $U=0.5$, $T=0.0274$ and system sizes $12^2, 16^2, 24^2$. The existence of two maximums shown by vertical arrows indicate the first-order phase transition, which weakens with increasing system size.}
\label{fig4}
\end{figure}

We also study the optical conductivity, $\sigma(\omega)$, and the Drude weight, $D$, extracted from a linear response of the current to an applied infinitesimal electric field\footnote{$\sigma'(\omega)$ is obtained in the second-order perturbative expansion of $\bs h[n_f]$ in applied flux and is averaged over all configurations $n_f$. It includes all vertex corrections}, and shown in Fig.~\ref{fig3}. 
Finally, Fig. \ref{fig4} depicts the f-electron properties (specific heat, staggered charge susceptibility) and the transition into the charge-ordered state by evaluating crossings of Binder cumulants \cite{Binder1981} and energy histograms \cite{Maska2006} at different $L^2$.

We start by introducing every phase in the full phase diagram, see Fig.~\ref{fig1}, of the model at half-filling:

(FG) At $U=0$, the model describes a trivial Fermi gas with  $\lim_{\omega\to 0} \text{DOS}(\omega)= \text{const}$, illustrated in Fig. \ref{fig2}(h). The real part of the conductivity $\sigma'(\omega)$ has unitary Drude weight, i.e. $\sigma'(\omega) = D \delta(\omega)$ with $D=1$. 
Fig. \ref{fig3}(f) shows that for any finite temperature $D=0$ for $U > 0$. As in the infinite dimensional case \cite{Si1992}, this phase is unstable for any $T$.

(WL/AI) is an interesting region overlooked in previous studies. The effect of the  $f$-electron averaging enters as a disorder potential for the $c$ species and induces an Anderson localization of the single particle eigenstates at low energies. 
Here, for sufficiently small $U$ and any finite $L$, the localization length becomes of the order of the volume. In both regions we observe a finite DOS at zero energy, i.e. $\lim_{\omega\to 0} \text{DOS}(\omega)\neq 0$, see Fig.~\ref{fig2}-(a,b).  
This is further corroborated by the results presented in Fig.~\ref{fig2} (h), where we performed the binning of the DOS within a fixed energy window around the Fermi level to avoid any ambiguity with the artificial broadening of $\delta$-peaks. At  $T<1$ and  $2\leq U \leq 7$, the $\text{DOS}$ acquires temperature dependence, which is attributed to non-local fluctuations due to proximity to the CDW phase \cite{Antipov2014, Ribic2016}.
The behavior of the IPR can be seen in Fig.~\ref{fig2}-(d-f) to differ  in the two regimes: in the WL throughout the spectrum we find $\text{IPR}$ strongly dependent on the system size,  $\text{IPR}(\omega)\propto V^{-1}$, except at the band edges,  signaling predominantly delocalized states. In the AI region, on the other hand, we find $\text{IPR}(\omega)\propto V^{0}$ for $\omega$ in a finite window around zero, $|\omega| < \Delta_{\text{ME}}$ marking the localized region of the spectrum. The low-energy states are localized. With further increase of $U$ this energy window expands, while the density of localized states decreases and approaches zero at the MI phase. 
Fig.~\ref{fig2} (g) shows that the $\mathrm{IPR}(0)$ as a function of $U$ has a $V^{-1}$ scaling at small $U$ and an approximate crossing  point at larger $U$. This crossing point is only weakly dependent on $L^2$ and was used to extract the approximate crossover line, shown in Fig.~\ref{fig1}. 
 
The conducting properties of WL/AI phase are studied in Fig.~\ref{fig3}, which shows the real part of the conductivity $\sigma'(\omega)$ as a function of $\omega$ at $T=0.5$. In the WL region (Fig.~\ref{fig3} (a,b)), $\sigma'$ is zero at $\omega = 0$, but the finite size scaling at small $\omega$ and $L \leq 48$ implies that  $\lim_{\omega\to 0^+} \sigma'(\omega) \to \text{const}$. In the AI region (Fig.~\ref{fig3}(c,d)), no such scaling takes place, and $\sigma'(0) = 0$. We observe $\sigma'(\omega) \propto   e^{-\Delta_\text{ME}/T} a(\omega)$ where $\Delta_\text{ME}$ is the energy gap between the Fermi level and the energy of the first delocalized state $\text{IPR}(\omega \lesssim \Delta_\text{ME} ) \propto V^{-1}$ and $a$ is roughly linear with frequency at $\omega \to 0$.
This temperature dependence is therefore compatible with that for activated hopping (Fig. \ref{fig3}e).

(MI) At high temperatures and sufficiently large $U$, a Mott-like phase sets in where the $c$-electrons develop a charge gap,  $\text{DOS}(\omega=0)=0$ for $\left| \omega\right|<\Delta$, while charge order is absent, see Fig.~\ref{fig2}(c). 
This phase is adiabatically connected to the point $t/U=0$ where $\Delta=U$ is the energy cost associated with occupying a $c$-electron site with $n_{f,\bs r}=1$. 

(CDW) For any non-zero interaction $U\neq0$  spontaneous symmetry breaking takes place as temperature is lowered below $T_{\mbox{\tiny CDW}}$ leading to a long-range checkerboard-ordered phase, see Ref. \cite{Maska2006} and Table \ref{table:FSS_exponents}. 
This phase is characterized by a zero-temperature gap around $\omega=0$, i.e. $\text{DOS}(\omega, T\to 0)=0$ for $\left| \omega\right|<\Delta$ with a charge gap $\Delta=U$. Within this phase, exact DMFT results for $d\to\infty$  reported a more complex internal structure with some sub-phases including sub-gap states at small $U$ \cite{Hassan2007,Matveev2008,Lemanski2014}.  

We now turn to a discussion of the nature of the transitions between different regions of phase space. 

\begin{table}[t]
	\centering
	\begin{tabular}{l*{4}{c|}}
		$U$    & $T_c$        & $\nu$           & $\gamma$         \\
		\hline
		3.0    &  $0.1370$    & $0.97 \pm 0.07$ & $1.76 \pm 0.02$ \\	
		5.0    &  $0.1339$   & $0.92 \pm 0.11$ & $1.74 \pm 0.07$ \\
		7.0    &  $0.1171$    & $1.00 \pm 0.11$ & $1.74 \pm 0.07$ \\
		10.0   &  $0.0951$    & $1.02 \pm 0.02$ & $1.73 \pm 0.07$ \\
		12.0   &  $0.0824$  & $1.02 \pm 0.17$ & $1.74 \pm 0.07$ \\
	\end{tabular}
	\caption{\footnotesize{Critical exponents $\gamma$ (susceptibility) and $\nu$ (correlation length) of the CDW transition for different values of $U$. The Ising exponents are $\gamma = 1.75$ and $\nu=1$.}}
	\label{table:FSS_exponents}
\end{table}

CDW transition - at large $U$ we find a previously reported transition between a disorder state at large $T$ and a CDW at small $T$ \cite{Kennedy1986}.   The transition between the two is of  Ising universality  with an order parameter given by the staggered $f$-occupation $\phi_\text{st} = \sum_{\bs r} e^{i \left\{\pi,\pi\right\}.\bs r}(2n_{f, \bs r}-1)$ as illustrated in Figs.~\ref{fig4}(a-b)  by the $T$-dependence of specific heat $C_v$ and $f$-electron susceptibility at momentum $Q=\left\{\pi,\pi\right\}$ for $U=6$. 
In fact, for large $U$, an exact mapping to the 2d-Ising model can explicitly be given \cite{Kennedy1986}. 
Numerically, the transition temperature is determined by the crossing of the Binder cumulant  $B_Q(L)$ \cite{Binder1981}, see Figs.~\ref{fig4}(c-d).
Upon decreasing $U$,  the high temperature disorder phase evolves from a MI to an AI, but the nature of the transition into the CDW state is maintained, as highlighted by the Ising exponents, see Table \ref{table:FSS_exponents}. 
In agreement with  previous studies  we find that for  $0  < U \lesssim 3$, the phase transition appears to be first order \cite{Maska2006}.
This is illustrated by the double peaked energy histogram in Fig.~\ref{fig4}e with maxima denoted by arrows. 
Interestingly, the disappearance of the double peak at $U\approx 3$  coincides with the WL-Al crossover of the high temperature phase.
The finite size scaling of Fig.~\ref{fig4}e shows that the first order nature of the transition weakens with increasing system size implying a  continuous transition in the infinite volume limit. 
This provides further evidence that the occurrence of the WL phase is a finite size effect. 
In the thermodynamic limit the AI phase extends until $U\to 0^+ $ and the transition into the charge-ordered state is continuous for all values of $U>0$.  

AI-MI transition - the existence of this transition in the FK model has not been reported previously. The MI is characterized by $\text{DOS}(0)=0$. As $U$ is reduced the onset of the AI phase can be best read off by $\text{DOS}(0) > 0$, as illustrated in Fig.~\ref{fig2}h, while $\sigma'(0)=0$ \footnote{We used the numerical criterion of $\text{DOS}(0) > 10^{-7}$ in our energy units of $t$}. 
As the $f$-electrons act as static scattering potentials the
grand canonical partition function of the FK model can be understood as that of annealed disorder. 
Across the transition  $f$-electron observables remain smooth. 
For asymptotically large temperatures FK maps to a binary random disorder model previously studied in the context of binary alloys 
 \cite{Dean1972,Kirkpatrick1972} where transitions similar to the AI-MI transition have previously been observed.  
Our results suggest that for finite temperature the universal features of the AI-MI transition are the same as the infinite temperature ones and the annealed nature of the disorder plays no qualitative role in this regime.  

WL-AI crossover - this crossover turns out to be a feature of finite system size. The crossover is reflected in the AC conductivity and is also clearly visible in the $\mbox{IPR}(\omega=0)$ which displays qualitatively different behavior in the WL, AI and MI phases.
As discussed above, at large temperatures, the nature of the disorder is not relevant as all possible configurations of the $f$ electrons are equally probable. Therefore, results resembling those of quenched disorder are to be expected. For 2D these imply the existence of a so-called weak localized regime where the localization length $\zeta\sim 1/\mbox{IPR}(\omega=0)$ depends on the coupling constant $U$ in an exponential fashion \cite{Abrahams1979}. Therefore, for any finite system of linear size $L$ and sufficiently small $U$, there is a regime where $L<\zeta$. In this case, the system displays properties of a bad metal. 

Having described the overall phase diagram of the two-dimensional half-filled FK model we now turn to a discussion of the significance of our findings. 
The fact that $\sigma'(\omega \to 0^+)\neq 0$ in the temperature range above the charge ordered state at small $U$, see Fig.~\ref{fig1}, suggests the identification of the WL regime with the previously reported metallic phase of the model~\cite{Freericks2003, Maska2006}. As shown above this regime is a feature of the finiteness of the underlying lattice and vanishes in the thermodynamic limit. Within the DMFT, the metallic phase originates from a finite local DOS$(\omega=0)$ and the neglect of spatial correlations~\cite{Freericks2003,Antipov2014}. The AI phase is captured by disorder extensions of the DMFT with the addition of quenched disorder  ~\cite{Byczuk2004,Tran2007,Gusmao2008,Byczuk2009,Byczuk2010}.

In this work, we primarily addressed the intricacies of the phase diagram above the charge ordering transition. Naively, the onset of charge order with propagation vector $Q$ in the weakly interacting or small $U$ regime is associated with the instability due to a  perfectly nested Fermi surface. Although DOS$(\omega=0)\neq 0$, the absence of a Fermi surface implies that even for arbitrarily small $U$ this CDW picture cannot apply, when coming out of the AI phase. It thus might be worthwhile to extend our analysis to the charge ordered part of the phase diagram.
Interestingly, as can be read off from Table~\ref{table:FSS_exponents}, the critical exponents associated with the onset of charge order in that region coincide with those of the classical two-dimensional Ising model. 
This raises the possibility that ordering transitions, traditionally interpreted within the Stoner theory of delocalized electrons, should be better described within a strong-coupling framework. This relates to the on-going debate between the weak- or strong-coupling nature of the onset of order near the emergence of superconductivity in {\it e.g.} the iron-based superconductors and the heavy fermions \cite{Si2008,Davis2013}.

Our results may also shed some light on the finite temperature phase diagram of the three-dimensional half-filled Hubbard model, where at low temperature an antiferromagnetic phase sets.
Ignoring the dynamic nature of the antiferromagnetic order parameter the system can be described  by a static model with annealed vector disorder. Thus, one may expect an Anderson localized phase at high temperatures separating the weak coupling metallic phase from the Mott insulator at large $U$.  
A recent study of the Anderson-Hubbard model with spin-dependent disorder leads to FK-like physics in parts of the phase diagram \cite{Skolimowski2016} to which our findings may be relevant.

Finally, we note that our results may be directly relevant to studies of localization in cold atoms \cite{Ospelkaus2006,Sanchez-Palencia2010,Fialko2009}. The model serves as a prototype for recent implementations of mass unbalanced fermions in optical lattices \cite{Jotzu2015, Greif2015,Ates2005} and the physical properties of the model extend past the infinite mass ratio regime \cite{Liu2015,Ueltschi2004}. It would be interesting to see an experimental confirmation of the existence a the localized phase for a translational-invariant system in the absence of explicit disorder. A direct verification with ultracold atoms systems should be possible with state-of-the-art technology. 

\begin{acknowledgments}
The authors acknowledge fruitful discussions with B.~Altshuler, E.~Castro, E.~Gull, E.~J.~K\"{o}nig, J.P.F.~LeBlanc, Y.~Bar~Lev, R.~M\"{o}ssner, Y.~Shchadilova, H.~Terletska, D.~Vollhardt, X.~Wan, and V.~Yudson. Our code has been developed using  TRIQS open source library (version 1.1) \cite{Parcollet2015}.
P.~Ribeiro acknowledges support by FCT through the Investigador FCT contract IF/00347/2014.
S.~Kirchner acknowledges partial support by the National Science Foundation of China, grant No.11474250, and the National Key R\&D Program of the MOST of China (No.2016YFA0300202).
\end{acknowledgments}

\bibliographystyle{apsrev_clean}
\bibliography{fkmcpaper}

\end{document}